\documentclass[draft_temp]{imsart}

\RequirePackage{amsthm,amsmath,amsfonts,amssymb}
\RequirePackage{natbib}
\RequirePackage[colorlinks,citecolor=blue,urlcolor=blue]{hyperref}
\RequirePackage{graphicx}

\usepackage{booktabs}
\usepackage{comment}
\usepackage{bbold}
\usepackage{ulem}
\usepackage{eurosym}

\usepackage{tikz}  
  
\usetikzlibrary{shapes, arrows, calc, arrows.meta, fit, positioning} 
\tikzset{  
    -Latex,auto,node distance =1.5 cm and 1.3 cm, thick,
    state/.style ={ellipse, draw, minimum width = 0.9 cm}, 
    point/.style = {circle, draw, inner sep=0.18cm, fill, node contents={}},  
    bidirected/.style={Latex-Latex,dashed}, 
    el/.style = {inner sep=2.5pt, align=right, sloped}  
}  

\startlocaldefs

\theoremstyle{remark}

\newtheorem*{example}{Example}

\endlocaldefs

 \defcitealias{Beis2020}{BEIS, 2020}
 \defcitealias{Beis2021}{BEIS, 2021b}
 \defcitealias{DecarbScheme}{BEIS, 2021}
 \defcitealias{UKCP18}{UKCP, 2018}
 \defcitealias{Northumberland}{NCC, 2018}
 \defcitealias{acola2021}{ACOLA, 2021}
 \defcitealias{EnergyPrices}{BEIS, 2022}

\newcommand{\bb}[1]{\boldsymbol{#1}}

\newcommand{\stanx}{\boldsymbol{\mathrm{x}}}

\begin{document}

\begin{frontmatter}
\title{A Bayesian decision support system in energy systems planning}
\runtitle{A Bayesian decision support system in energy systems planning}

\begin{aug}
\author[A]{\fnms{Victoria} \snm{Volodina}\ead[label=e1]{v.volodina@ucl.ac.uk}},  
\author[B]{\fnms{Nikki} \snm{Sonenberg}\ead[label=e2]{}},
\author[C]{\fnms{Peter} \snm{Challenor}\ead[label=e3]{}}
\and 
\author[D]{\fnms{Jim Q.} \snm{Smith}\ead[label=e4]{}}.

\address[A]{Clinical Operational Research
Unit, University College of London, London, United Kingdom,
\printead{e1}}

\address[B]{Heilbronn Institute
for Mathematical Research, University of Bristol, Bristol, United Kingdom,
\printead{e2}}

\address[C]{College of Engineering, Mathematics and Physical Sciences, University of Exeter, Exeter,  United Kingdom,
\printead{e3}}

\address[D]{Department of Statistics, University of Warwick, Coventry, United Kingdom.
\printead{e4}}

\end{aug}

\begin{abstract}

Gaussian Process (GP) emulators are widely used to approximate complex computer model behaviour across the input space. Motivated by the problem of coupling computer models, recently progress has been made in the theory of the analysis of networks of connected GP emulators. In this paper, we combine these recent methodological advances with classical state-space models to construct a Bayesian decision support system. This approach gives a coherent probability model that produces predictions with the measure of uncertainty in terms of two first moments and enables the propagation of uncertainty from individual decision components.

This methodology is used to produce a decision support tool for a UK county council considering low carbon technologies to transform its infrastructure to reach a net-zero carbon target. In particular, we demonstrate how to couple information from an energy model, a heating demand model, and gas and electricity price time-series to quantitatively assess the impact on operational costs of various policy choices and changes in the energy market.

 \end{abstract}

\begin{keyword}
\kwd{Gaussian process}
\kwd{state-space model}
\kwd{decision support}
\end{keyword}

\end{frontmatter}

\section{Introduction}
The current state of the energy sector is considered as a key contributor to climate change due to its high dependence on fossil fuel. Various   programs have been proposed to enable energy transition, a pathway towards net zero emission energy systems, such as the Energy White Paper in UK \citep{WhitePaper2020}. 
At a regional level in UK, local governments are faced with the challenge of implementing energy efficiency improvements to achieve a net-zero carbon state, for which there are available funding and grants, for example, the Public Sector Decarbonisation Scheme (\citetalias{DecarbScheme}).
Despite undeniable reductions in carbon emissions associated with green technologies, there are a number of risks that need to be reviewed by councils, who are responsible for the day-to-day oversight over energy generation, energy consumption, emissions capture, policy and engagement. 
Factors that influence these risks include the employment of green technologies that are less mature than the conventional ones \citep{Newbery2016}, as well as short-term volatility of energy prices \citep{Liu2021}.
Here, we present a decision support tool that can inform a local council about immediate effects of changes in policy and energy markets on energy infrastructure as well as producing cost projections crucial for short-term planning.

Energy systems planning is complex since it is characterised by multiple interdependent processes and relies on various sources of information. Complex computational models are increasingly deployed as part of decision support for system operation, planning and policy  \citep{Hall2016}, and time series data of gas and electricity prices provide another valuable source of information \citep{Barons2021}. 
To construct the decision support system, a feed-forward graph is used
to describe how one variable can influence another and draw together computer models' simulations and time-series data to inform various variables of interest.

For variables informed by deterministic computer models, we specify a Gaussian Process (GP) emulator to provide a statistical representation of the model. Conditioned on a set of computer model evaluations, a GP emulator is used to generate predictions together with a measure of uncertainty about the predicted model output at inputs that have not been tested.
GP emulators have been widely used in climate and environmental studies \citep{Conti2009,williamson2014evolving,volodina2020diagnostics}, energy electricity prices  \citep{Wilson2022} and transport infrastructure \citep{svalova2021emulating}.  

For variables associated with time-series data,  dynamic linear models (DLMs), an important class of state-space models, are used to provide a description of the time-varying relationship between a collection of independent variables and a response \citep{West2006, Petris2009}. In addition, the computation is straightforward as the Kalman filter provides us with the closed-form expressions for estimation and forecasting. 

To construct a composite model that can provide a trustworthy representation of the whole system, we are required to integrate statistical models for individual variables, which can be computationally cumbersome for large systems. Since stakeholders and decision-makers are mainly interested in predictions (expectation) and the uncertainty measure about the prediction (variance), 
we iteratively use the law of total expectation and the law of total variance to obtain the projections for the variable of interest.
Our approach resembles Bayes Linear methods which use expectation as a primitive to express and manage uncertainty by recognising the difficulties involved in specifying a full probability distribution \citep{Goldstein2007}.  
The proposed framework allows us to explicitly account for the uncertainty linked to individual variables in our projections of the variable of interest, an important property for decision-makers and stakeholders \citep{Aqua2015}.

The work in this paper operationalises recent methodological advances in the theory of networks of connected GP emulators used to analyse systems of computer models 
\citep{Kyzyurova2018,Sanson2019,Ming2021}.
In addition, we adopt results about the multiregression dynamic model (MDM) \citep{Queen1993, Queen2008, Queen2009} to incorporate time-series data in the composite model. Motivation of this approach has similarities with the integrating decision support systems (IDSS) of \cite{Leonelli2015} and  \cite{Leonelli2020}.

 The rest of this paper is organised as follows, in the remainder of this section we present the case study on energy systems planning. In Section \ref{sec:prelim} we define the individual component models used in the construction of the composite model in Section \ref{sec:linked}. 
 We present the results for the case study in Section \ref{sec:application}. Concluding remarks are given in Section \ref{sec:conc}.

\subsection{Case study}\label{subsec:casestudy}
The proposed methodology has been developed as part of the project\footnote{This work was funded by the EPSRC National Centre for Energy Systems Integration through the flex fund award FFC3-008.} ``An integrating decision support system in energy planning.'' The aim of the project was to develop  a demonstrator decision support system to generate projections of operational costs associated with a mix of heating technologies to meet the projected heating demand as well as to study the changes in the projections under various policies and events in the energy market. During this project, we partnered with a local county council, that considered the introduction of an electricity powered ground source heat pump (GSHP) for a council facility (leisure centre), where currently a seventeen year old gas boiler is used to generate heat.

The cost of setting up the GSHP would be covered by a grant from the Public Decarbonisation Scheme (\citetalias{DecarbScheme}), a state initiative to fund heat decarbonisation and energy efficiency measures. However, the council is still responsible for day-to-day expenses (operational costs) associated with heating the leisure centre. Given that the price of electricity is currently higher compared to gas, the new heating technology can potentially drive up the operational costs and introduce extra risks related to the volatility in the electricity price market.

\begin{figure}[h!]
\begin{center}
\begin{tikzpicture}  
    \node[state] (a) at (0,0) {  $Y_2$}; 
     
    \node[state] (b) [right =of a] { $Y_3$};  
    \node[state] (d) [below =of a] { $Y_1$};  

    \node[state] (c) [below =of b] {$Y_4$};  

    \path (a) edge  (b); 
     \path (d) edge (c);  
   \path (a) edge (c);  
      \draw (b) -- (c);  
\end{tikzpicture} 
\end{center}
     \caption{A feed-forward graph for energy planning, with heating demand $Y_1$,  gas price $Y_2$, electricity price $Y_3$, and  operational cost $Y_4$.} 
    \label{fig:EnergyDAG}
\end{figure}
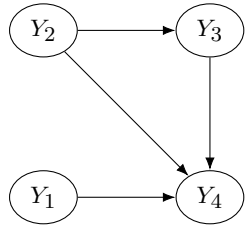

Following discussions with the experts in the council, a simplified version of a network model for energy systems planning was agreed, and is shown in Figure \ref{fig:EnergyDAG}. We identified three important factors influencing operational costs $(Y_4)$: heating demand ($Y_1$), gas price $(Y_2)$ and electricity price $(Y_3)$. The dependence of electricity price on gas price is included, since natural gas is one of the main fuels used to generate electricity \citep{Mosquera2019}. It is assumed that there is no dependence between the heating demand and energy prices as we are considering the heating demand in a building that largely depends on the indoor temperature, outdoor temperature and the thermal characteristics of the building \citep{Larsen2020}.

Heating demand ($Y_1$) and operational costs ($Y_4$) are generated by two computational models. To generate the heating demand we use a heating building energy demand model, which relies on a degree days approach \citep{DeRosa2014}.  Degree days data is a simplified representation of outside temperature and widely used to determine changes in heating and cooling demands \citep{Spinoni2018}.   
 We consider input parameters including efficiency of the equipment, global building transmission coefficient, the domains of which were elicited from the council including the half-hourly data of the energy (kWh) used from 2017-2021 by the leisure centre and internal reports. The operational costs are derived from the energy model, based on OSeMOSYS \citep{howells2011} an open source model, which computes the energy supply mix, in terms of generation capacity and energy delivery. The variables $Y_2$ and $Y_3$ represent natural gas price and electricity price, respectively, and publicly available data (\citetalias{EnergyPrices}) is used to model these variables.

In this paper, for the purposes of demonstration, we primarily use open source models and publicly available data. In practice we expect that decision makers would have access to more information, e.g., confidential data and to proprietary software used by industry and government.

\section{Preliminaries}\label{sec:prelim}

We define the individual components of the system model, the Gaussian process model in Section \ref{sec:GP_model} and dynamic linear model for time series analysis in Section \ref{subsec:statespace}.

\subsection{Gaussian Process model}
\label{sec:GP_model}

Let $\stanx=(x_1, \dots, x_p)\in\mathbb{R}^p$ be a vector of inputs and $f(\stanx)$ be the scalar-valued output that represents the process of interest. We define a statistical model for $f(\stanx)$ as a sum of three processes
\begin{equation}
    f(\stanx)=\bb{h}(\stanx)^T\bb{\beta}+\epsilon(\stanx)+\nu(\stanx),
\end{equation}
where $\bb{h}(\stanx)^T\bb{\beta}$ represents the global response surface behaviour, $\epsilon(\stanx)$ is a correlated residual process capturing local input dependent deviation from the global response surface modelled as a zero-mean Gaussian process with covariance function $k(\stanx, \stanx'; \sigma^2, \bb{\delta})=\sigma^2r(\stanx, \stanx'; \bb{\delta})$  and $\nu(\stanx)$ is a nugget process representing the noise in the response defined as a zero-mean Normal with variance $\tau^2$. We specify a small value for $\tau^2$ to improve the numerical stability
\citep{Andrianakis2012}.

The probability function for $f(\stanx)$ conditioned on the statistical model parameters $\{\bb{\beta}, \sigma^2, \bb{\delta}, \tau^2 \}$ is
\begin{equation}
\label{eq:prior_dist}
    f(\stanx)\vert \bb{\beta}, \sigma^2, \bb{\delta}, \tau^2\sim\text{GP}\Big(\bb{h}(\stanx)^T\bb{\beta}, k(\cdot, \cdot; \sigma^2, \bb{\delta}, \tau^2) \Big),
\end{equation}
with  $\bb{h}(\stanx)=[h_1(\stanx), \dots, h_q(\stanx)]^T$, unknown regression coefficients $\bb{\beta}=(\beta_1, \dots, \beta_q)^T$, and  covariance 
\begin{align}\label{eqn_temp_label}
k(\stanx, \stanx'; \sigma^2, \bb{\delta}, \eta)=\sigma^2r(\stanx, \stanx'; \bb{\delta})+\tau^2\mathbb{1}\{\stanx = \stanx' \},
\end{align}
where $\sigma^2$ and $\tau^2$ are variance parameters. We choose the squared exponential correlation function $r(\stanx, \stanx'; \bb{\delta})=\exp\Big\{-\sum_{i=1}^p\Big(\frac{x_i-x_i'}{\delta_i} \Big)^2 \Big\}$, with correlation length parameters $\delta_i>0$, with $\bb{\delta}=(\delta_1, \dots, \delta_p)$ and where $\mathbb{1}\{\cdot\}$ is the indicator function. 

Suppose we observe $n$ model realisations $\mathrm{F}= (f(\stanx_1), \dots, f(\stanx_n) )$ at design points $\mathrm{X}= (\stanx_1, \dots, \stanx_n )$.  
Under the assumption of a non-informative prior for $\bb{\beta}$ and $\sigma^2$, $\pi(\bb{\beta}, \sigma^2)\propto 1/\sigma^2$,
a posterior for $f$ can be found, conditional on $\{\mathrm{X}, \mathrm{F}\}$ and $\bb{\delta}, \tau^2$ is the Student Process with $n-p$ degrees of freedom with mean
\begin{equation}
    \label{eq:post_mean2}
    \mu^*(\stanx^*)=\bb{h}(\stanx^*)^T\hat{\bb{\beta}} + \bb{r}(\stanx^*)\bb{R}^{-1}\big(\mathrm{F}-\bb{H}(\mathrm{X})\hat{\bb{\beta}} \big),
\end{equation}
and variance
\begin{align}
    \label{eq:post_var2}
    \sigma^{*2}(\stanx^*, \stanx^*)= \hat{\sigma}^2&\Big(1+\tau^2-\bb{r}(\stanx^*)\bb{R}^{-1}\bb{r}(\stanx^*)+\big(\bb{h}(\stanx^*)-\bb{H}(\mathrm{X})^T\bb{R}^{-1}\bb{r}(\stanx^*) \big)^T\\
    &\times \big(\bb{H}(\mathrm{X})^T\bb{R}^{-1}\bb{H}(\mathrm{X}) \big)^{-1}\big(\bb{h}(\stanx^*)-\bb{H}(\mathrm{X})^T\bb{R}^{-1}\bb{r}(\stanx^*) \big)\Big),
\end{align}
where 
\begin{align}
\hat{\bb{\beta}}=\big(\bb{H}(\mathrm{X})^T\bb{R}^{-1}\bb{H}(\mathrm{X})\big)^{-1}\bb{H}(\mathrm{X})^T\bb{R}^{-1}\mathrm{F},
\end{align}
and 
\begin{align}
\hat{\sigma}^2=\frac{\mathrm{F}^T\big(\bb{R}^{-1}-\bb{R}^{-1}\bb{H}(\mathrm{X})(\bb{H}(\mathrm{X})^T\bb{R}^{-1}\bb{H}(\mathrm{X}))^{-1}\bb{H}(\mathrm{X})^T\bb{R}^{-1}\big)\mathrm{F}}{n-p-2}
\end{align}
where $\bb{r}(\stanx^*)$ is an $n$-vector whose $i^{\text{th}}$ component is $r(\stanx^*, \stanx_i)$, the correlation between the point of interest $\stanx^*$ and the design point $\stanx_i$, where $\bb{H}(\mathrm{X})=[\bb{h}(\stanx_1), \dots, \bb{h}(\stanx_n)]^T$ and $\bb{R}$ is an $n\times n$ correlation matrix with entries $\bb{R}_{ij}=r(\stanx_i, \stanx_j)$. 
We use \texttt{RobustGaSP} package in $\mathrm{R}$ to perform optimisation of the marginal posterior function in order to find maximum a posterior (MAP) values of $\bb{\delta}$ and $\tau^2$  \citep{RobustGP2020}. 

\subsection{Dynamic linear model (DLM)}
\label{subsec:statespace}
Consider a time-series $(Y(t), t=1, 2, \dots, n)$, where $Y(t)$ denotes the series value observed at time $t$. For each $t$, the model is defined by observation and system equations, 
\begin{equation}
\label{eq:DLM_def}
  \begin{split}
    Y(t)&= \bb{F}(t)^T\bb{\theta}(t)+v(t), \quad v(t)\sim N(0, V(t)), \\
    \bb{\theta}(t)&=\bb{G}(t)\bb{\theta}(t-1) + \bb{w}(t), \quad \bb{w}(t)\sim N(\bb{0}, \bb{W}(t)),
\end{split}
 \end{equation}
where $\bb{G}(t)$ and $\bb{F}(t)$ are known and have dimensions ($p\times p$) and ($p\times 1$), respectively.
The error term $(v_t)$ and error vector $(\bb{w}_t)$ are internally and mutually independent.  The recurrence relationships for sequential updating of posterior distributions have closed form expressions and are equivalent to the Kalman filter \citep{West2006}.

Given initial prior information $\mathcal{D}_0$ at $t=0$, we specify 
$        \bb{\theta}(0)\vert \mathcal{D}_0\sim N(\bb{m}(0), \bb{C}(0))$
and denote the available information set at  time $t$ by $\mathcal{D}_t=\{Y(t), \mathcal{D}_{t-1}\}$. At  $t-1$, the posterior is  
$ (\bb{\theta}(t-1)|\mathcal{D}_{t-1})\sim N(\bb{m}(t-1), \bb{C}(t-1))$,
with prior 
$     (\bb{\theta}(t)|\mathcal{D}_{t-1})\sim N(\bb{a}(t), \bb{R}(t))$,
    where $\bb{a}(t)=\bb{G}(t)\bb{m}(t-1)$ and $\bb{R}(t)=\bb{G}(t)\bb{C}(t-1)\bb{G}(t)^T+\bb{W}(t)$.
The one-step ahead forecast is
    \begin{align}
    (Y(t)\vert \mathcal{D}_{t-1})\sim \text{N}(f(t), Q(t)),
    \end{align}
    where $f(t)=\bb{F}(t)^T\bb{a}(t)$ and $Q(t)=\bb{F}(t)^T\bb{R}(t)\bb{F}(t)+V(t)$,
with posterior  
        \begin{align}
        (\bb{\theta}(t)\vert\mathcal{D}_t)\sim N(\bb{m}(t), \bb{C}(t)),
    \end{align}
    where  $\bb{m}(t)=\bb{a}(t)+\bb{A}(t)e(t)$ and $\bb{C}(t)=\bb{R}(t)-\bb{A}(t)Q(t)\bb{A}(t)^T$ with $e(t)=Y(t)-f(t)$ and $\bb{A}(t)=\bb{R}(t)\bb{F}(t)Q(t)^{-1}$.
 
 For $k\geq 1$, the $k$-step marginal state distributions are
        \begin{align}
        (\bb{\theta}(t+k)|\mathcal{D}_t)\sim N(\bb{a}(t+k), \bb{R}(t+k)),
    \end{align}
with forecast distribution
        \begin{align}
        (Y(t+k)|\mathcal{D}_{t})\sim N(f(t+k), Q(t+k)), 
        \end{align}
    where $f(t+k)=\bb{F}(t+k)^T\bb{a}(t+k)$ and $Q(t+k)=\bb{F}(t+k)^T\bb{R}(t+k)\bb{F}(t+k)+V(t)$,
    that can be recursively calculated using
$        \bb{a}(t+k)=\bb{G}(t+k)\bb{a}(t+k-1)$, and 
$        \bb{R}(t+k)=\bb{G}(t+k)\bb{R}(t+k-1)\bb{G}(t+k)^T+\bb{W}(t+k)$.

We assume constant variance and define precision parameters $\psi_1=V^{-1}$ and $\psi_2=W^{-1}$,  assumed to be a priori independent and specify
$\psi_i\sim\mathcal{G}(3, 0.01), \quad i=1, 2$, commonly used values.
We use \texttt{RStan}, \texttt{R} interface to \texttt{Stan} \citep{Carpenter2017}
to implement Markov Chain Monte Carlo and find MAP estimates for $\psi_1$ and $\psi_2$. An alternative approach is to allow $V$ to vary stochastically by specifying $\phi_t=1/V_t$ as part of the discounted variance learning \citep{West2006}.

\section{The probabilistic integrating structure}
\label{sec:linked}
We present results on the coupling of time-series models in Section \ref{sec:link_ts} and linked GP emulators in Section \ref{sec:prob_int}.

\subsection{Linked time-series models}
\label{sec:link_ts}

Let $\bb{Y}(t)=(Y_1(t), \dots, Y_n(t))^T$ denote an $n$-dimensional time-series,   
where $Y_r(t)$ corresponds to the observation of component $r$ at time $t$ and define $\bb{Y}_r^t=(Y_r(1), Y_r(2), \dots, Y_r(t))^T$ and $\bb{Y}^t=(\bb{Y}_1^t,\bb{Y}_2^t, \dots, \bb{Y}_n^t)^T$. 

Suppose that the variables are ordered and indexed so that there is a conditional independence structure related to causality, then for $t\in\mathbb{N}$ and $r=2, \dots, n,$
\begin{align}
   Y_r(t)&\perp\{ \{Y_1(t), \dots, Y_{r-1}(t) \}\setminus\text{pa}(Y_r(t))\}|\text{pa}(Y_r(t)),\\
   Y_r(t)&\perp\{ \{\bb{Y}_1^t, \dots, \bb{Y}_{r-1}^t \}\setminus\text{pa}(\bb{Y}_r^t)\}| \ \text{pa}(\bb{Y}_r^t), \bb{Y}_r^{t-1},
\end{align}
where the set $\text{pa}(Y_r(t))$ is called a parent of $Y_r(t)$. 
We model $Y_1(t)$ and the conditional probability models $Y_r(t)|\text{pa}(Y_r(t))$, $r=2, \dots n$ as DLMs, so that forecasts for these models can be obtained using the results in Section \ref{subsec:statespace}.
We proceed to define a Multiregression Dynamic Model (MDM) for $\bb{Y}(t)$, a class of multivariate state-space time-series models that preserve the conditional independence relationships between variables over time \citep{Queen1993, Queen2009}. 
As the inference for the MDM can be performed by breaking down the multivariate model into univariate DLMs, there are computational  advantages.
 
Since $Y_r(t)$ and $\text{pa}(Y_r(t))$ are observed simultaneously, we require the marginal forecast distributions for $Y_r(t), r=2, \dots, n$.  In general, the marginal forecast distributions for $Y_r(t), r=2, \dots, n$ will not have simple form, however there exist closed-form expressions for mean and variance, sufficient for forecasting purposes.
In particular, we can obtain the closed-form expressions for mean and variance of the marginal forecast distribution for $Y_r(t)$ iteratively,
\begin{align*}
    \mathbb{E}[Y_r(t)|\mathcal{D}_{t-1}]&=\mathbb{E}\Big[\mathbb{E}(Y_r(t)|\text{pa}(Y_r(t)), \mathcal{D}_{t-1})|\mathcal{D}_{t-1} \Big],\\
    \mathbb{V}[Y_r(t)|\mathcal{D}_{t-1}]&=\mathbb{E}\Big[\mathbb{V}(Y_r(t)|\text{pa}(Y_r(t)), \mathcal{D}_{t-1})|\mathcal{D}_{t-1} \Big]+\mathbb{V}\Big[\mathbb{E}(Y_r(t)|\text{pa}(Y_r(t)), \mathcal{D}_{t-1})|\mathcal{D}_{t-1} \Big].
\end{align*}

\subsection{Linked GP emulators}
\label{sec:prob_int}

\begin{figure}[b!]
\begin{center}
\begin{tikzpicture}  
    \node[draw=none,fill=none] (a) at (0,0) {  $\stanx_1$};

      \node[draw=none,fill=none] (c) [below =of a] {  $\stanx_2$};

        \node[state] (b) [right =of a] { $Y_1$}; 
        
            \node[state] (d) [right =of c] { $Y_2$}; 
            
            \node[state] (e) [right =of d] { $Y_r$};  

              \node[draw=none,fill=none] (f) [below =of c] {  $\stanx_{r-1}$};  

            \node[state] (g) [right =of f] { $Y_{r-1}$}; 

              \node[draw=none,fill=none] (h) [below =of e] {  $\stanx_r$};

    \path (a) edge (b);
    \path (b) edge (e);  
    \path (c) edge (d);  
    \path (d) edge (e);  
    \path (f) edge (g);  
    \path (g) edge (e);  
    \path (h) edge (e);

\end{tikzpicture} 
\end{center}
\caption{A two-layered computer system, where $Y_1, \dots, Y_r$ represent the responses of $r$ computer models, and $\stanx_1, \dots, \stanx_r$ correspond to exogenous inputs.} 
\label{fig:LinkedGPs}
\end{figure}
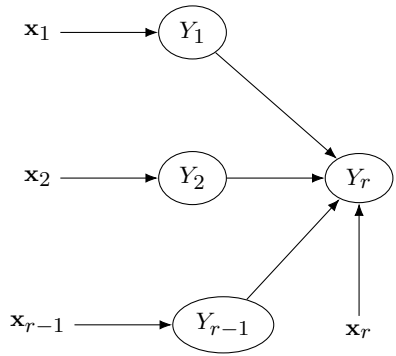

We present a method for linking GP models, originally introduced to emulate a system of computer simulators, where output of one computer simulator is an input to another \citep{Kyzyurova2018}). Consider a system, as illustrated in Figure \ref{fig:LinkedGPs}, where
$Y_1, \dots, Y_{r-1}\in \text{pa}(Y_r)$, produce outputs that feed into  $Y_r$. We model each computer simulator's output as a stationary GP defined in Section \ref{sec:GP_model}:
\begin{align*}
    Y_i|\stanx_i&\sim \text{GP}(\mu_i^*(\cdot), \sigma_i^{*2}(\cdot, \cdot)), 
    \end{align*}
  where $Y_i\in\text{pa}(Y_r)$, and 
    \begin{align*}
    Y_r|\text{pa}(Y_r), \stanx_r&\sim\text{GP}(\mu_r^{*}(\cdot), \sigma_r^{*2}(\cdot, \cdot)).
\end{align*}
Given the global (exogenous) inputs $\stanx_1, \stanx_2, \dots, \stanx_r$, the predictive distribution 
is 
\begin{equation}
\label{eq:integral}
    p(Y_r|\stanx_1, \dots, \stanx_r)=\int_{\text{pa}(Y_r)}p(Y_r|\text{pa}(Y_r), \stanx_r)p(\text{pa}(Y_r)|\stanx_1, \dots, \stanx_{r-1})d\text{pa}(Y_r).
\end{equation}
However, $(Y_r\vert \stanx_1, \dots, \stanx_r)$ is neither analytically tractable nor Gaussian in general. \cite{Sanson2019} computed the first two moments numerically using Monte Carlo samples under the assumption that the densities in Equation (\ref{eq:integral}) are Gaussian.
Under the normality assumption together with specification of squared exponential correlation function for individual GP emulators, \cite{Kyzyurova2018}   provided the close-form analytical expressions for mean and variance of linked emulator, and \cite{Ming2021} extended these expressions under a class of Mat\'{e}rn correlation functions. Similar analytical results were presented for a system of two computer models by \cite{Marque2019} under some mild conditions.

To construct a linked emulator, we adopt the assumptions introduced by \cite{Ming2021}, that is, the regression function in the GP for the computer model $Y_r$ is linear and a squared exponential kernel is specified for all models. It is also assumed that 
$\text{pa}(Y_r)\sim\text{MVN}(\bb{\mu}, \bb{\Sigma})$,
where $\bb{\mu}=[\mu_1^*(\stanx_1), \dots, \mu_{r-1}^*(\stanx_{r-1})]$ and $\bb{\Sigma}$ is the covariance matrix with $\text{diag}(\bb{\Sigma}) = \big( \sigma_1^{*2}(\stanx_1), \dots, \sigma_{r-1}^{*2}(\stanx_{r-1})\big ).$ Then the linked GP is defined as a normal approximation to $p(Y_r|\stanx_1, \dots, \stanx_r)$ with analytical mean and variance given by
\begin{align}
    \mu_L^*(\stanx_1, \dots, \stanx_r)&=\mathbb{E}\big[\mathbb{E}[Y_r|\text{pa}(Y_r), \stanx_r] \big],\\
    \sigma_L^{*2}(\stanx_1, \dots, \stanx_r)&=\mathbb{E}\big[\mathbb{V}[Y_r|\text{pa}(Y_r),\stanx_r]\big]+\mathbb{V}\big[\mathbb{E}[Y_r|\text{pa}(Y_r),\stanx_r]\big].
\end{align}
Since the presented approaches for linking time-series models and GP emulators are based on the law of total expectation and the law of total variance for computing mean and variance for the marginal posterior distribution of the quantity of interest, we claim these methods can be combined to construct a composite model, and present results for the energy systems application in Section \ref{sec:linkedgpmodels}.

\begin{example} 
Consider a system with two computer models composed sequentially, as illustrated in Figure \ref{fig:SystemPlot}. The computer models $f_1$ and $f_2$ with scalar-valued output $y_1$ and $y_2$ are defined as 
$$
y_1 = f_1(x)=2x\sin(x), \quad y_2 = f_2(y_1)=y_1^2\cos(y_1).
$$

We construct GP models for $f_1$ and $f_2$ with $h(x)=(1, x, x^2)$ and $h(y_1) = (y_1)$ and squared exponential correlation function. We use \texttt{RobustGaSP} (as above) to estimate individual GP models and fix $\bb{\delta}$ and $\tau^2$ at maximum a posterior (MAP) values.

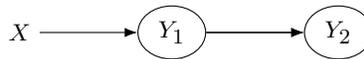
\begin{figure}[h!]
\begin{center}
\begin{tikzpicture}  
    \node[draw=none,fill=none] (a) at (0,0) {  $X$}; 
  
    \node[state] (b) [right =of a] { $Y_1$};  
    \node[state] (c) [right =of b] { $Y_2$};  

    \path (a) edge (b);
    \path (b) edge (c);  
 
     \draw (b) -- (c);  
\end{tikzpicture} 
\end{center}
\caption{Computer system in Example 3 where $f_1$ and $f_2$ have 1-D input and output.} 
\label{fig:SystemPlot}
\end{figure}

The predictions produced by a linked GP are illustrated in the right panel of Figure \ref{fig:UncertaintyPropagation} that show the mean and the two standard deviation prediction interval computed using the analytical expressions for mean and variance of linked emulator (RMSE=9.24). We observe that a linked emulator fails to predict the behaviour of $Y_2$ in the region where $3.5<x<4.5$,  as there are not enough design points to learn about the model behaviour in this region.
 
For comparison, we also consider modelling the relationship between $Y_2$ and $X$ directly and ignoring $Y_1$. The prediction and two standard deviation prediction intervals for this model are presented in the left panel plot of Figure \ref{fig:UncertaintyPropagation} (RMSE=14.57). We observe the discrepancy between the true and predicted values and wide prediction intervals, which imply greater degree of uncertainty about the response behaviour. The importance of explicitly including $Y_1$ is demonstrated, and illustrates the crucial lesson of uncertainty handling in decision support of energy systems planning. 

\begin{figure}[h!]
\begin{center}
\includegraphics[width = .91\textwidth]{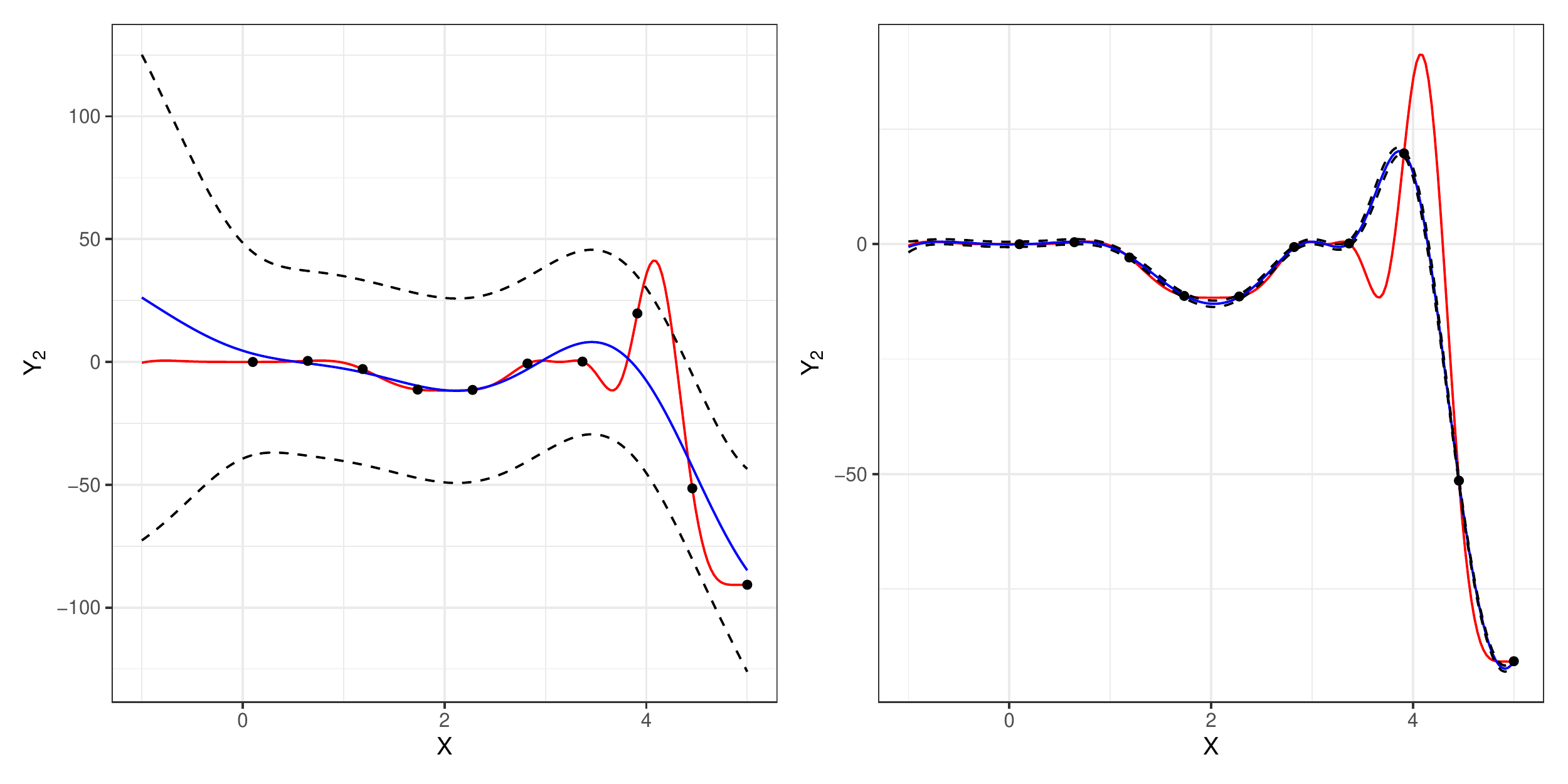}
\end{center} 
\caption{
\textit{Left panel plot}: a statistical model considering the relationship between $X$ and $Y_2$ and ignoring $Y_1$. \textit{Right panel plot}: predictions produced by a linked emulator. The red solid line is the true functional form between the input $X$ and output of the system; the blue solid line is the mean prediction; the dashed lines represent the two standard deviation prediction intervals; the black dots correspond to the design points.} 
\label{fig:UncertaintyPropagation}
\end{figure}
\end{example}

\section{Application}\label{sec:application}
We proceed to construct the decision support system for the case study in Section \ref{subsec:casestudy}.  The fitting of probability models for the individual variables is presented in Sections \ref{subsec:ts_energy} and  \ref{subsec:comp_models}. Then in Section \ref{sec:linkedgpmodels} we present the process of obtaining a composite model and produce the summaries of interest to decision-makers. 
Using the composite model, various scenarios are studied in Section \ref{sec:results}.
 
\subsection{Time-series model for energy prices, $Y_2(t), Y_3(t)$}\label{subsec:ts_energy}

\begin{figure}[h!]
\begin{center}
\includegraphics[width = 1\textwidth]{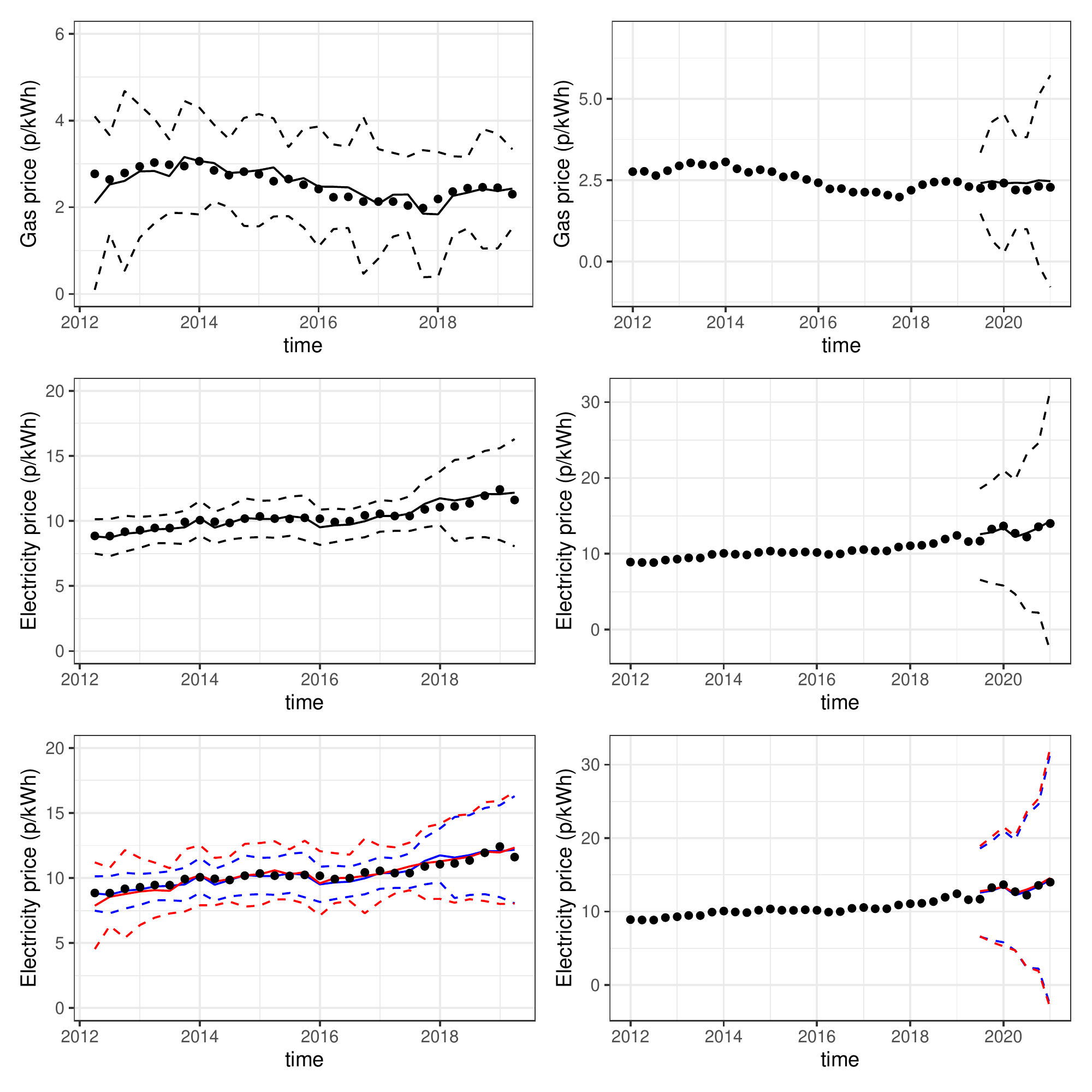}
\caption{Top row: dynamic linear regression fit for gas prices. Middle row: dynamic linear regression fit for electricity prices. Bottom row: MDM fit (red) and DLM fit (blue) for electricity prices. The solid line and dashed lines correspond to mean and two standard deviation prediction intervals. The points correspond to the observed prices.}\label{fig:Diagnostics_DLM}
\end{center} 
\end{figure}

 In this model it is assumed that natural gas prices are a function of market supply and demand, thus gross gas production (GWh), net imports (GWh) and levels of natural gas stored (GWh) are considered as major supply-side factors affecting the price, whereas the price of coal (p/kWh) as a demand-side factor \citep{Brown2008}. These factors are reported in the form of a quarterly time series from 2012 to 2021 (\citetalias{EnergyPrices}). We specify the form of the dynamic linear model for gas prices as:
 \begin{align}
     Y_2(t)&=\bb{F}_2(t)^T\bb{\theta}_2(t) + v_2(t), \quad v_2(t)\sim N(0, V_2), \\
     \bb{\theta}_2(t)&=\bb{G}_2(t)\bb{\theta}_2(t-1)+\bb{w}_2(t), \quad \bb{w}_2(t)\sim N(\bb{0}, \bb{W}_2),
 \end{align}
 where $\bb{F}_2(t)^T=(1, \text{Prod}_t, \text{Imports}_t, \text{Storage}_t, \text{Coal}_t)$ and $\bb{G}_2(t)$ is $5\times 5$ identity matrix.
We note that for fitting purposes the  variables $\text{Prod}$, $\text{Imports}$ and $\text{Storage}$ are scaled by $1e^{-5}$. To validate the performance of DLM, we produce the one-step-ahead forecasting and $k$-step ahead forecasting in the first row of Figure \ref{fig:Diagnostics_DLM}. The predictions are close to the observed values, whereas observed values lie within two standard deviation prediction interval.

As with natural gas prices, the electricity price dynamic depends on a large number of factors. Traditional drivers of electricity prices are natural gas and coal prices, since these two types of fuels are mainly used to generate electricity \citep{Mosquera2019}. Another important supply factor is the availability of renewable energy sources such as wind and sun. Electricity markets have become more dependent on these sources of energy due to the policies promoting sustainable energy generation \citep{Mulder2013}. Further, incentive schemes or reducing fossil-fuel electricity generation, such as carbon prices have a significant effect on electricity price dynamic \citep{Aatola2013}.

Based on the above and discussions with the council, the following factors were chosen as explanatory variables of dynamic regression model: gas prices (p/kWh), Emissions Trading System (ETS) price (\euro/\ tCO2) and the share of electricity generated by offshore wind (\%). Similar to gas prices, we consider the quarterly values for electricity price factors  \citepalias{EnergyPrices}. We model electricity prices as
\begin{align}
    Y_3(t)&=\bb{F}_3(t)^T\bb{\theta}_3(t)+v_3(t), \quad v_3(t)\sim N(0, V_3), \\
    \bb{\theta}_3(t)&=\bb{G}_3(t)\bb{\theta}_3(t-1)+\bb{w}_3(t), \quad \bb{w}_3(t)\sim N(\bb{0}, \bb{W}_3),
\end{align}
where $\bb{F}_3(t)^T=(1, \text{Gas}_t, \text{ETS}_t, \text{OffshoreWind}_t)$ and $\bb{G}_3(t)$ is $4\times 4$ identity matrix. To validate the performance of DLM, we produce the one-step-ahead forecasts and $k$-step ahead forecasts in the second row of Figure \ref{fig:Diagnostics_DLM}.

\subsection{GP emulators for computational models, $Y_1$, $Y_4$}
\label{subsec:comp_models}

There are two types of heating demand: baseline demand, that is present throughout the year, and seasonal demand. Based on half-hourly energy data provided by the council, we approximate the baseload energy consumption. 

Let $Y_1$ correspond to the quarterly seasonal heating demand produced by the heating demand model, described in Section \ref{subsec:casestudy}. We study the three inputs: surface temperature, efficiency of the equipment and the global building transmission coefficient, since we believe that these inputs are associated with a high level of uncertainty. In addition, these inputs could  be considered as part of the decision-makers policies and scenario analysis.

Table \ref{table:domain_demand} provides the domains of input parameters considered for emulation. To learn about the behaviour of model output in response to changes in inputs, we produce a  space-filling design, that is, we use a 100-run maximin distance Latin Hypercube (LHC) to generate a model ensemble \citep{Morris1995}. We use the first 80 points as a design (training set) and retain the remaining points for validation.

We construct a GP emulator using the method in Section \ref{sec:GP_model}, specifying a linear form of the regression function, $h(\stanx)=(1, x_1, x_2, x_3)$, where $x_1=\text{number of HDD}$ (heating degree days), $x_2=\text{efficiency}$ and $x_3=\text{transmission coefficient}$, a standard form that includes a constant and linear terms in each component of $\stanx$. 
We also adopt a squared exponential correlation function and use \texttt{RobustGaSP} to fit the individual GP models.

\begin{figure}[h!]
\begin{center}
\includegraphics[width = .8\textwidth]{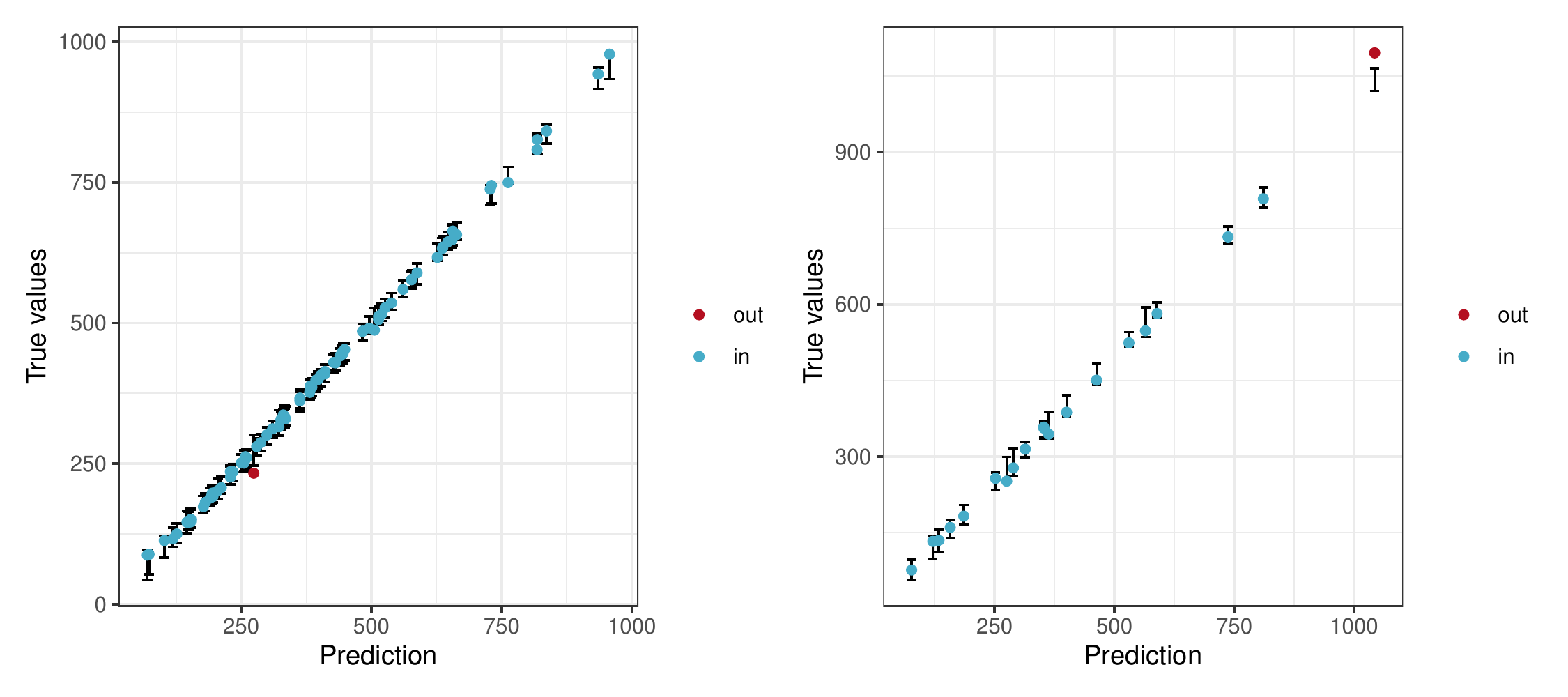}
\end{center} 
\caption{Traditional Leave One Out (\textit{left panel}) and cross-validation (\textit{right panel}) plots for heat demand model emulator.  The predictions and two standard deviation prediction intervals are in black. The true values are in either green, if they lie within two standard deviations of the predictions, or red otherwise.}
\label{fig:LOOGPEnergyDemand}
\end{figure}

\begin{table}[h!]
\caption{Input parameters for the emulation of the heat demand model}
\centering
\begin{tabular}{ll} 
 \toprule
 \textbf{Input parameter (unit)} & \textbf{Domain} \\
 \midrule
 Number of HDD in a quarter  & $[200, \ 1200]$\\
 Efficiency of the equipment, & $[0.5, \ 1]$\\
 Global building transmission coefficient, $H$ & $[10, \ 25]$\\
 \bottomrule
\end{tabular}
\label{table:domain_demand}
\end{table}
To validate the performance of the GP emulator, we produce the traditional leave one out and cross-validation plots in Figure \ref{fig:LOOGPEnergyDemand}.
We plot the emulated values and the observed function output on the $x$-axis and $y$-axis, respectively, together with the error bars corresponding to two standard deviation prediction intervals. The points are either coloured green if they are within the two standard deviation prediction intervals, or red otherwise. We observe that the emulator predictions lie close to the true values and the size of the error bars are small, 
and then conclude that the emulator is performing well.

\begin{table}[h!]
\caption{Input parameters for the emulation of the energy model}
\centering
\begin{tabular}{ll} 
 \toprule
 \textbf{Input parameter (unit)} & \textbf{Domain} \\
 \midrule
Quarterly heating demand (kWh)   & $[48,000,   \ 1,450,000]$\\
 Gas price (p/kWh) & $[1, \  5]$\\
 Electricity price (p/kWh) & $[4, \ 25]$\\
 Efficiency of gas boiler & $[0.3, \  1]$\\
 Efficiency of heat pump & $[2, \  6]$\\
 \bottomrule
\end{tabular}
\label{table:domain_energy}
\end{table}

To construct a GP emulator for energy system model, let $Y_4$ denote the quarterly operating costs (£) generated by the energy system model described in Section \ref{subsec:casestudy}. The energy system model takes various inputs such as heating demand, fuel prices and efficiency of heat technologies, and Table \ref{table:domain_energy} provides the domains of these input parameters considered for emulation. We proceed to provide details of the statistical emulator used to model the quarterly operating costs as a function of heating demand, gas and electricity prices, and efficiencies of gas boiler and heat pump.

We generate a \textit{space-filling} design, i.e., a 160-run maximin distance Latin Hypercube (LHC) to produce an  ensemble. We chose the first 120 points as a design and retain the remaining points for validations. To construct a GP emulator, we specified a linear form of the regression function $h(\stanx)=(1, x_1, x_2, x_3, x_4, x_5)$, where $x_1=\text{heating demand}$, $x_2=\text{gas price}$, $x_3=\text{electricity price}$, $x_4=\text{efficiency of gas boiler}$ and $x_5=\text{efficiency of heat pump}$.

\begin{figure}[h!]
\begin{center}
\includegraphics[width = .8\textwidth]{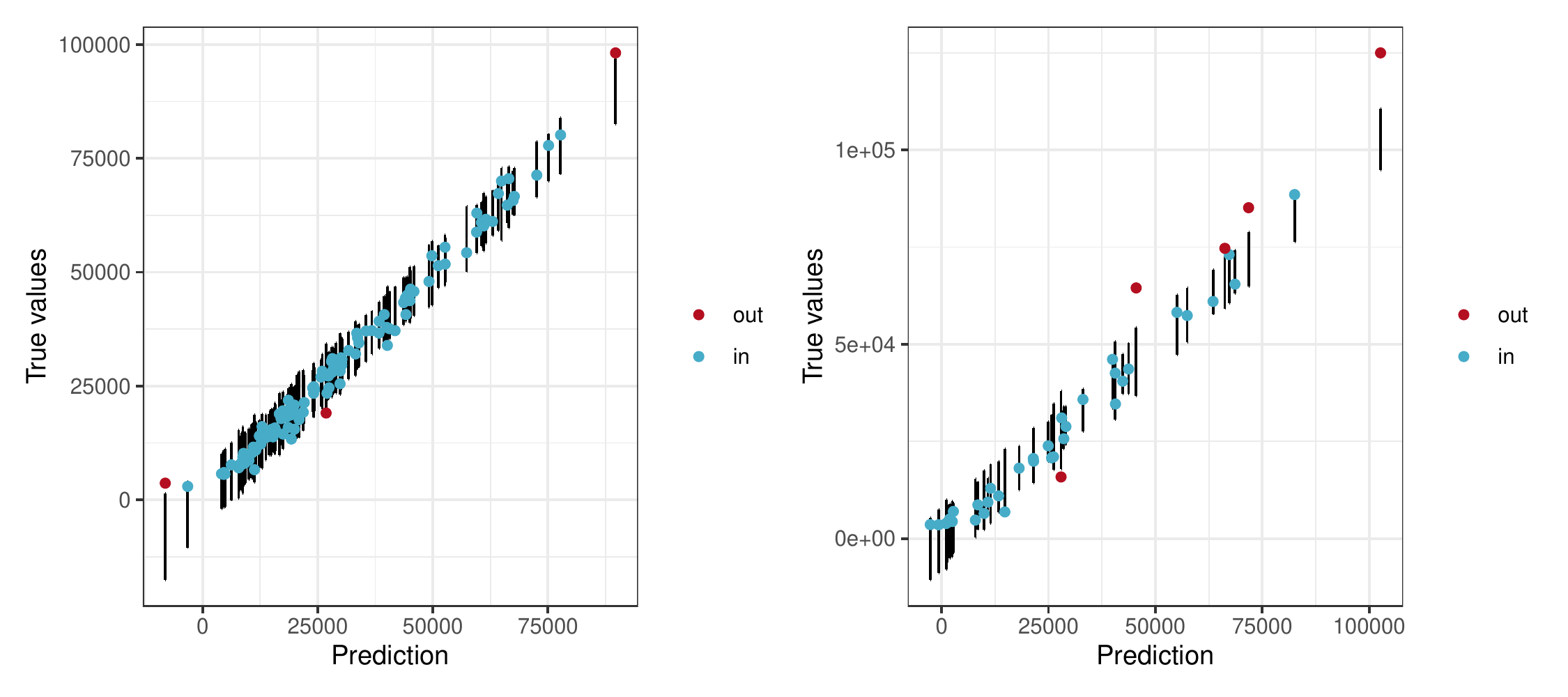}
\end{center} 
\caption{Traditional Leave One Out (\textit{left panel}) and cross-validation (\textit{right panel}) plots for energy systems model emulator.  The predictions and two standard deviation prediction intervals are in black. The true values are in either green, if they lie within two standard deviations of the predictions, or red otherwise.} 
\label{fig:LOOGPEnergyModel}
\end{figure}

To validate the performance of the GP emulator, we produce the traditional leave one out and cross-validation plots in Figure \ref{fig:LOOGPEnergyModel}. 
The plots indicate that the emulator still represents the energy model well, even though there are wider error bars compared to the energy demand model.

\subsection{Linking models}\label{sec:linkedgpmodels}
To obtain the mean and variance of the posterior predictive distribution of $Y_4$ using the linked emulator methodology (Ming and Guillas, \citeyear{Ming2021}), we require the mean and variance of the posterior predictive distributions of $Y_1$, $Y_2(t)$ and $Y_3(t)$. The DLM for $Y_3(t)$ defined in Section \ref{subsec:ts_energy}, is conditional on $Y_2(t)$, and we proceed to find the predictive mean and variance of the marginal distribution using the results from Section \ref{sec:link_ts}.

We start by computing the expectation and variance of the one-step ahead forecast distribution for $Y_2(t)$:
\begin{align}
    \mathbb{E}[Y_2(t)|\mathcal{D}_{t-1}, \stanx_2] &= f_2(t)=\bb{F}_2(t)^T\bb{a}_2(t), \\
    \mathbb{V}[Y_2(t)|\mathcal{D}_{t-1}, \stanx_2] &=Q_2(t)=\bb{F}_2(t)^TR_2(t)\bb{F}_2(t)+V_2(t),
\end{align}
where $\stanx_2$ is a vector exogenous (global) inputs considered as part of $Y_2(t)$ model.

The conditional forecast mean for $Y_3(t)$ is given by
\begin{align}
    \mathbb{E}[Y_3(t)|\mathcal{D}_{t-1}, y_2(t), \stanx_3]=f_3(t)=\bb{F}_3(t)^T\bb{a}_3(t),
\end{align}
where $\stanx_3$ is a vector of exogenous (global) inputs considered as part of $Y_3(t)$ model. Then
the mean for the forecast distribution for $Y_3(t)$ is
\begin{align}
    \mathbb{E}[Y_3(t)|\mathcal{D}_{t-1}, \stanx_3]&=\mathbb{E}[\bb{F}_3(t)\vert\mathcal{D}_{t-1}, \stanx_3]\bb{a}_3(t)=\tilde{\bb{\mu}}^T\bb{a}_3(t),
\end{align}
where $\tilde{\bb{\mu}}^T=(1, f_2(t), \stanx_3)$.

The conditional forecast variance for $Y_3(t)$ is given by
\begin{align}
    \mathbb{V}[Y_3(t)|\mathcal{D}_{t-1}, y_2(t), \stanx_3]&=Q_3(t)=\bb{F}_3(t)^TR_3(t)\bb{F}_3(t)+V_3(t),
\end{align}
 and we can compute the variance for the marginal forecast distribution for $Y_3(t)$ by
\begin{align}
\mathbb{V}[Y_3(t)|\mathcal{D}_{t-1}, \stanx_3]&=\mathbb{E}\big[\mathbb{V}[Y_3(t)|\mathcal{D}_{t-1}, y_2(t), \stanx_3] \big]+\mathbb{V}\big[\mathbb{E}[Y_3(t)|\mathcal{D}_{t-1}, y_2(t), \stanx_3] \big],\\
&=\text{tr}\{R_3(t)(\tilde{\bb{\mu}}\tilde{\bb{\mu}}^T+\tilde{\bb{\Omega}}) \} + \bb{a}_3(t)^T\tilde{\bb{\Omega}}\bb{a}_3(t) + V_3(t) = Q_3^*(t),
\end{align}
where $\tilde{\bb{\Omega}}=\text{diag}(0, Q_2(t), 0, 0)$. 
To compute the marginal forecast covariances between gas and electricity prices, $\mathbb{C}[Y_2(t), Y_3(t)|\mathcal{D}_{t-1}]$, by
Queen et al. (Theorem 1, \citeyear{Queen2008}), we obtain
\begin{align}
    \mathbb{C}(Y_2(t), Y_3(t)|\mathcal{D}_{t-1}, \stanx_2, \stanx_3)
    &=\mathbb{E}\big[Y_2(t)\times\mathbb{E}[Y_3(t)|\mathcal{D}_{t-1}, y_2(t), \stanx_3] \big] \nonumber\\
    &-\mathbb{E}[Y_2(t)|\mathcal{D}_{t-1}, \stanx_2]\mathbb{E}[Y_3(t)|\mathcal{D}_{t-1}, \stanx_3]\nonumber\\
    &=a_{13}(t)Q_2(t).
\end{align}
Similarly, we can obtain the mean and variance for the \textit{k-step}-ahead forecast distribution for $Y_3(t+k), k\geq 1$, by considering $\mathbb{E}[Y_2(t+k)|\mathcal{D}_t, \stanx_2]=f_2(t+k)$ and $\mathbb{V}[Y_2(t+k)|\mathcal{D}_t, \stanx_2]=Q_2(t+k)$ together with $\bb{a}_3(t+k)$ and $Q_3(t+k)$. 

We present the one-step-ahead and $k$-step ahead forecasts produced by the DLM (in blue) and MDM (in red) in the third row of Figure \ref{fig:Diagnostics_DLM}. We observe wider prediction intervals obtained with MDM compared to the individual DLM in particular for one-step-ahead, which indicates that we propagated the uncertainty about the gas price and accounted for this uncertainty in the predictions of the electricity price model.

After obtaining the first two moments of the marginal posterior distributions for the parents of $Y_4$, i.e., $\text{pa}(Y_4)=(Y_1, Y_2(t), Y_3(t))$, we  assume that
\begin{align}
    \text{pa}(Y_4)\sim MVN(\bb{\mu}, \bb{\Sigma}),
\end{align}
where $\bb{\mu}=[\mu_1^*(\stanx_1), f_2(t), \tilde{\bb{\mu}}^T\bb{a}_3(t)]$ and \begin{align}\bb{\Sigma}=\begin{pmatrix}
\sigma_1^{*2}(\stanx_1) & 0 & 0\\
0 & Q_2(t) & a_{13}(t)Q_2(t)\\
0 & a_{13}(t)Q_2(t) & Q_3^*(t)
\end{pmatrix}.
\end{align}
We then use Ming and Guillas (Theorem SM2.1, \citeyear{Ming2021})  to compute the mean and variance of the marginal posterior distribution for $Y_4$. We note that the time-varying components of $\bb{\mu}$ and $\bb{\Sigma}$ can be replaced by mean, variance and covariance for the $k$-step ahead forecast distributions.

\subsection{Results}\label{sec:results}
After performing model fittings, we considered three scenarios. Scenario 1 is where energy prices are kept at the same values reported by BEIS. Scenario 2 accounts for a 25\% increase in gas and electricity prices, and 
Scenario 3 represents a 30\% increase in electricity prices, and a sudden increase of 65\% in gas prices related to a 40\% increase in imported gas and a 50\% fall in stored gas.  
This third scenario is used to capture the current energy crisis in Europe \citep{Ambrose2021,Liu2021}.

\begin{figure}[b!]
\begin{center}
\includegraphics[width = .81\textwidth]{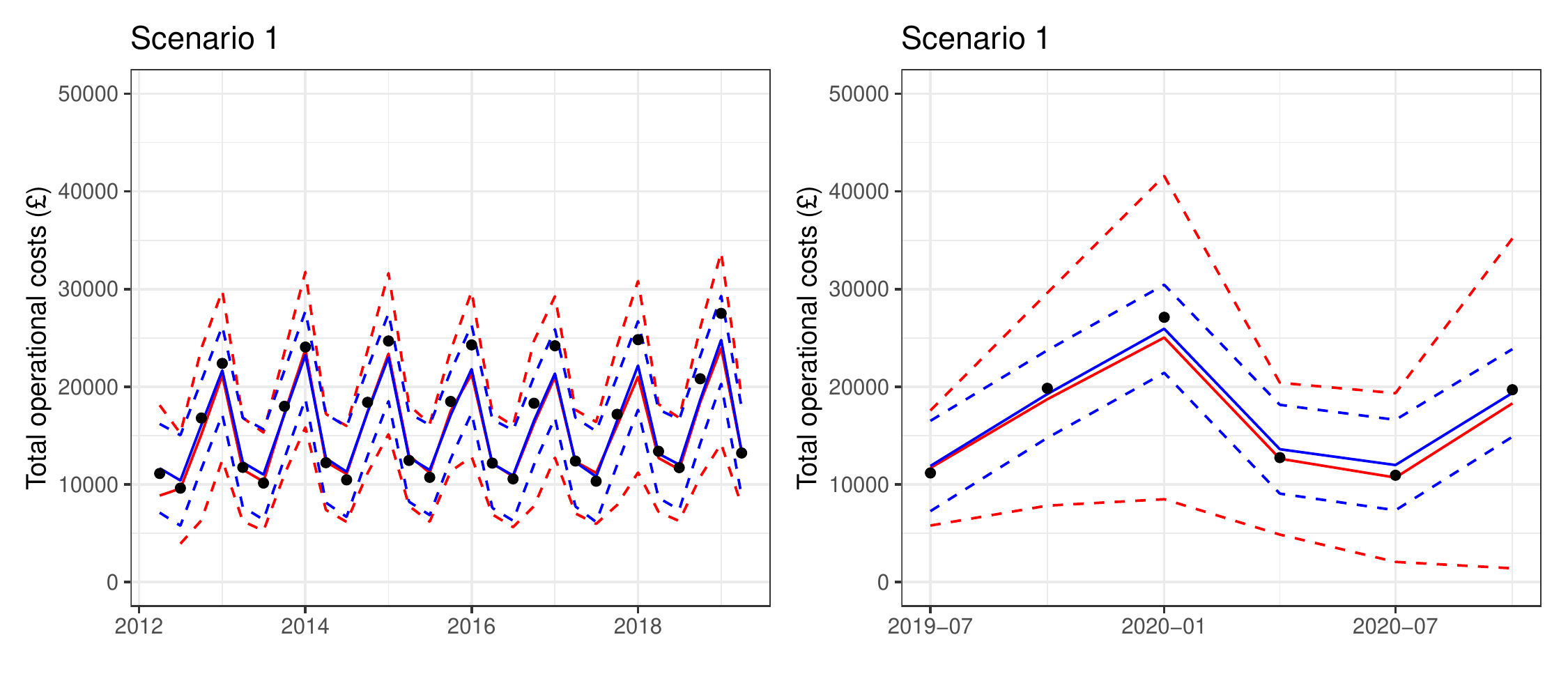}
\includegraphics[width = .81\textwidth]{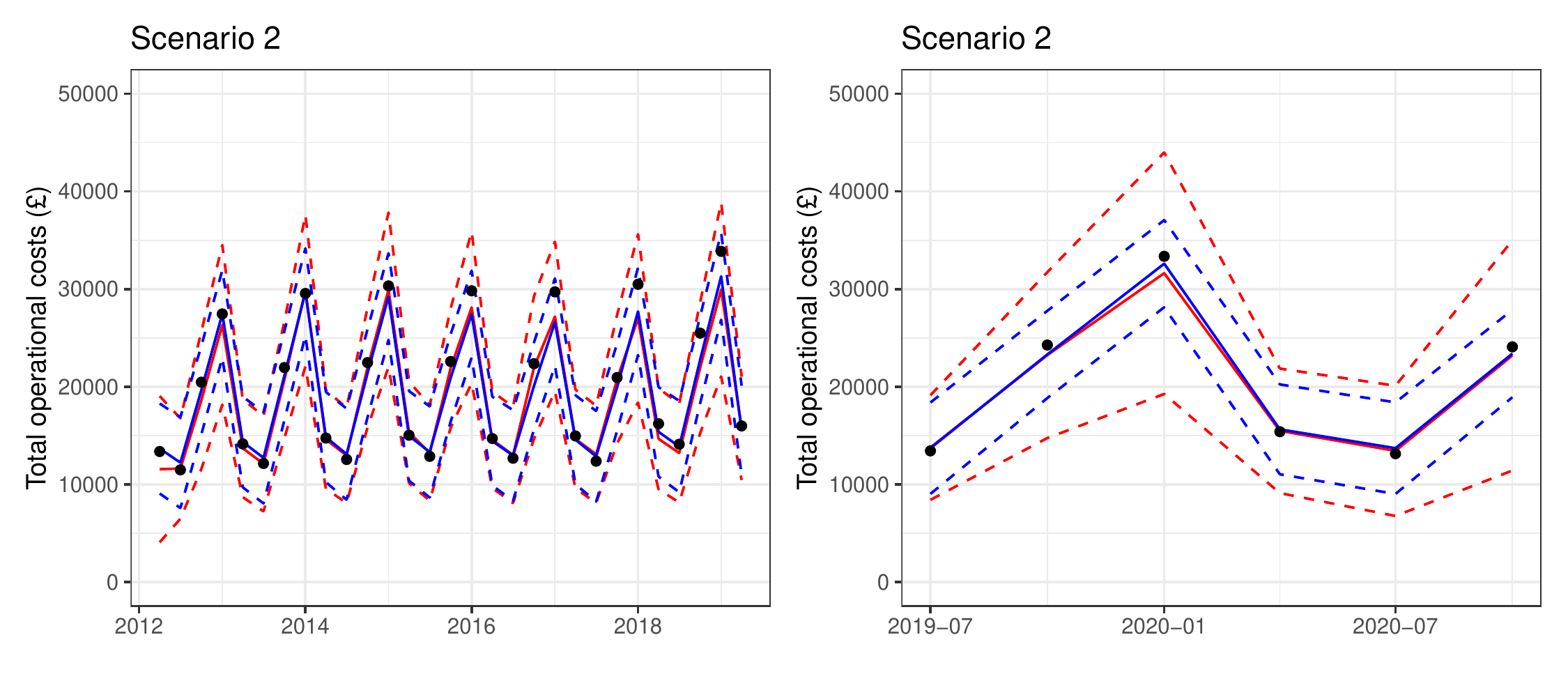}
\includegraphics[width = .81\textwidth]{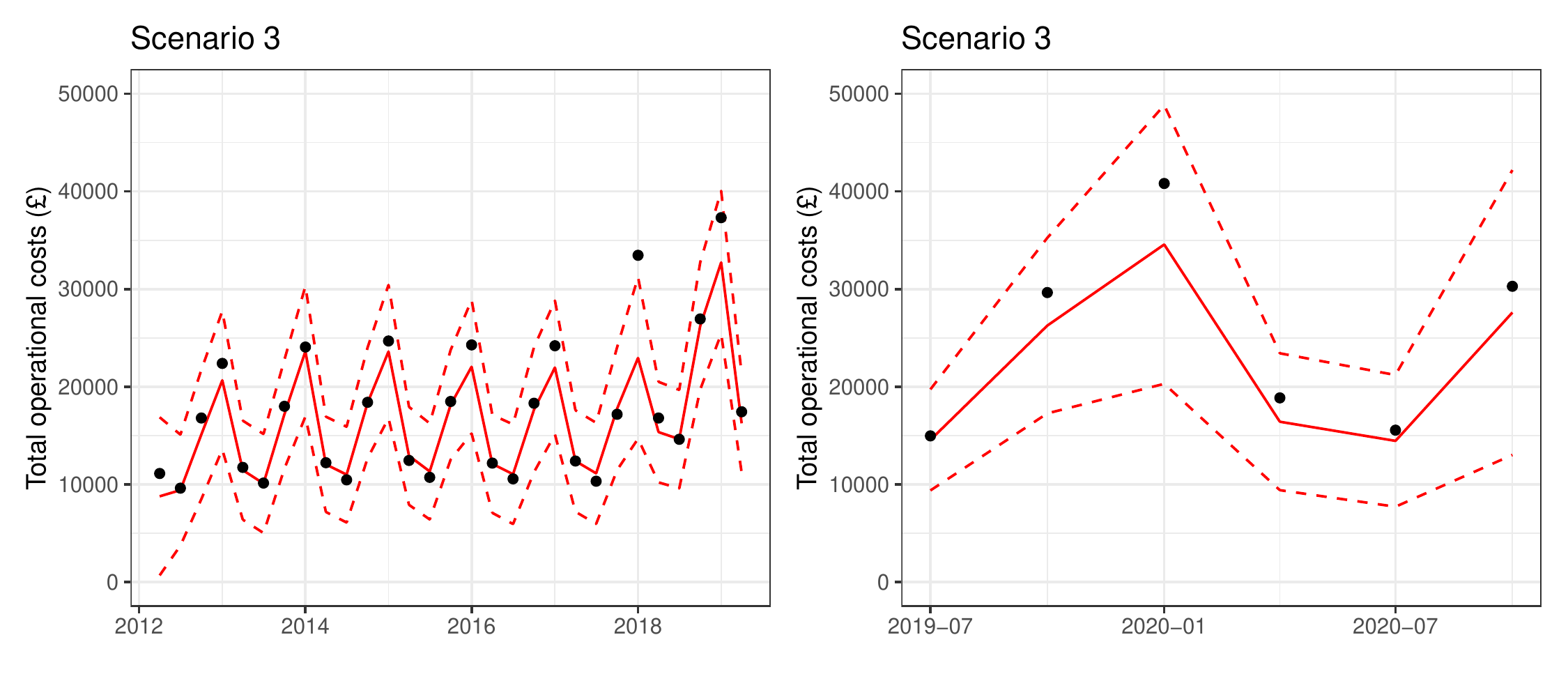}
\caption{One-step ahead predictions (right column) and forecasts (left column) for three scenarios under consideration. The blue line and blue dotted lines correspond to mean and two standard deviation prediction interval produced by an emulator for energy system model. The red line and the red dotted lines corresponds to mean and standard deviation prediction interval produced by a composite model. The points correspond to the observed simulations.  }\label{fig:filterfore_plot_linked_cases}
\end{center} 
\end{figure}

In Figure \ref{fig:filterfore_plot_linked_cases} we illustrate the effects of the changes in electricity and gas prices on operational cost projections as produced by the composite model (in red). For comparison, we also produce predictions generated by an emulator for the energy system model (in blue). From the first and second rows of Figure \ref{fig:filterfore_plot_linked_cases}, we observe wider prediction intervals produced by the composite model compared to an emulator for energy system model, indicating the propagation of uncertainty relating to individual factors, i.e., heating demand and energy prices, into the uncertainty about the operational costs. This property of the composite model is  crucial for decision-making process, since by not considering the uncertainty of the upstream variables, the prediction intervals fail to account for elements we know to be uncertain, leading to less informed decisions.

The predictions generated by the two statistical models are close to the observed values. We note that the one-step ahead predictions tend to underestimate high values of operational costs, and this needs to be further investigated. For Scenario 2, we observe an increase in operational cost projections due to a rise in energy prices. However, the prediction intervals produced by the composite model are narrower compared to Scenario 1. This is due to the fact that the gas price becomes comparatively cheaper than electricity price, and the energy system model chooses to rely on the gas boiler for heating with GSHP for back-up. Since there is less uncertainty associated with the gas price projections, this in turn translates into less uncertainty about the operational costs generated by the composite model. Further, Scenario 3 is used to demonstrate the ability of a composite model to adapt to price shocks in the energy market.

\section{Concluding remarks}\label{sec:conc}
Simulator models together with time-series data are valuable sources of information for decision-makers. By employing recent developments on networks of GP emulators and classical time-space models, we have demonstrated how to link together simulator models and time-series into a decision support system for an energy systems planning application.

The developed method has been applied specifically to a planning question proposed for a facility managed by a UK county council considering the use of green technologies and the effect of this decision on their operational costs. For three scenarios, we have produced projections, predictive mean and variance for total operational costs, explicitly taking into account uncertainties associated with various decision components identified by the council. We believe that the framework presented  can support planning decisions in complex environments and be extended to other decision support challenges in planning and policy such as clinical applications that rely on a wide scope of patient data generated by monitors and sophisticated healthcare models \citep{Hose2019}. 

There are a number of technical extensions to the proposed approach. Firstly, we are currently considering to adapt the experimental design criterion proposed by \cite{Ming2021}
to perform variance-based sensitivity analysis \citep{Saltelli2008} to measure the effect of individual variables or a collection of variables on a variable of interest within the decision support system. Further work involves weakening the strict assumptions introduced to obtain closed-form expressions for mean and variance of linked emulators such as specifying non-linear regression functions $h(\cdot)$ or nonstationary correlation functions.

There are a number of possible extensions to the work considering the application domain, for example, studying the simulation of storage in this context, in particular, whether this can reduce the uncertainty of the projections. Similar to \cite{Leonelli2015} and \cite{Barons2021}, we could further elicit the form
of the utility function from the panel of experts and compute formally expected utility scores for
different policies.

\begin{acks}[Acknowledgments]
 We would like to thank Chris Dent (University of Edinburgh) and Mark Roberts (Northumberland County Council) for their constructive and substantive comments.  
\end{acks}




\bibliographystyle{imsart-nameyear}  
\bibliography{library}
\clearpage
\begin{supplement}
\stitle{
}
 \sdescription{
 }
 Diagnostic figures are presented for Scenario 2 and Scenario 3 in Section \ref{sec:results}.

\begin{figure}[h]
\begin{center}
\includegraphics[width = 1\textwidth]{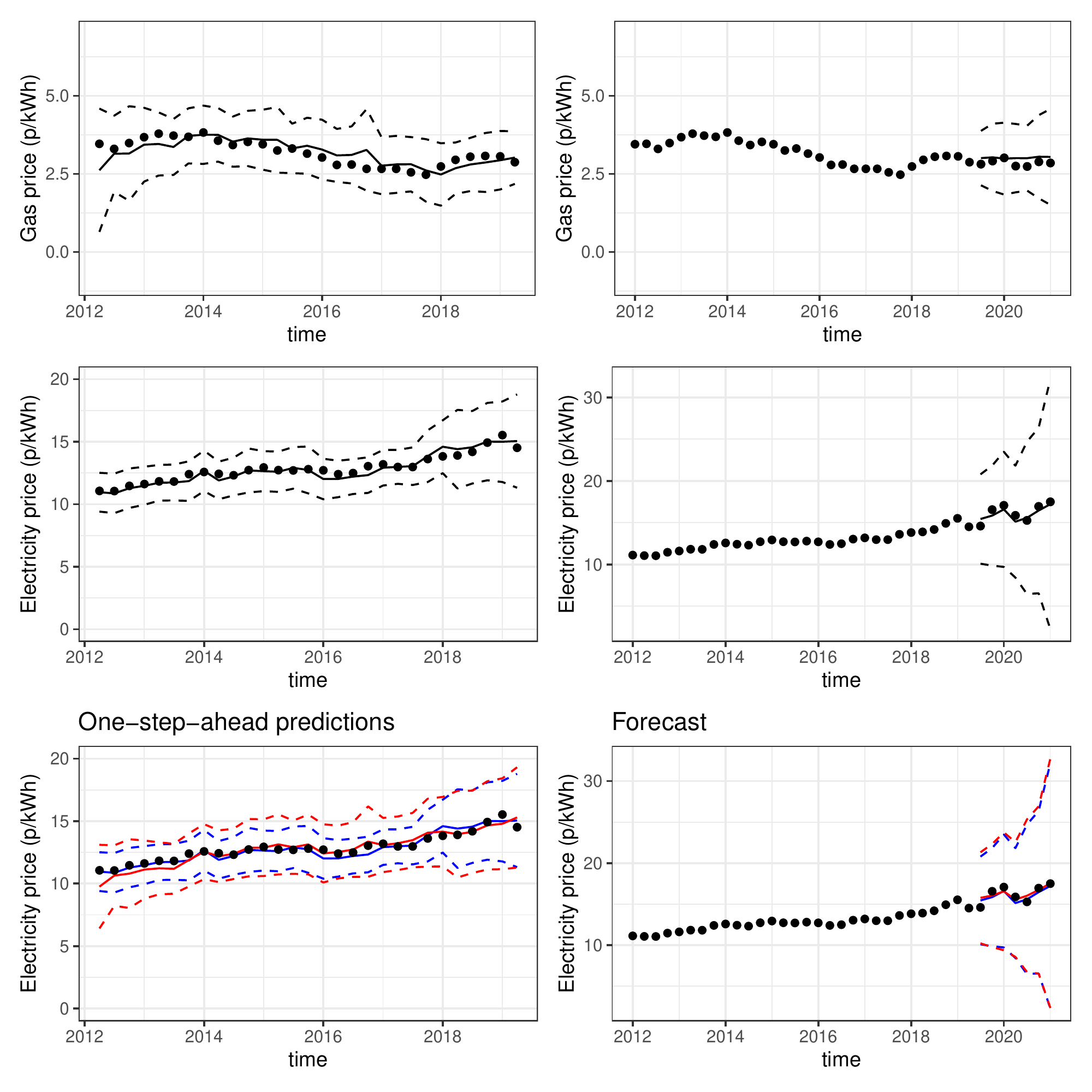}
\caption{Top row: dynamic linear regression fit for gas prices under Scenario 2. Middle row: dynamic linear regression fit for electricity prices under Scenario 2. Bottom row: LMDM fit (red) and DLM fit (blue) for electricity prices under Scenario 2. The solid line and dashed lines correspond to mean and two standard deviation prediction intervals. The points correspond to the observed prices.}\label{fig:Diagnostics_DLM_case2}
\end{center} 
\end{figure}

\begin{figure}[h!]
\begin{center}
\includegraphics[width = 1\textwidth]{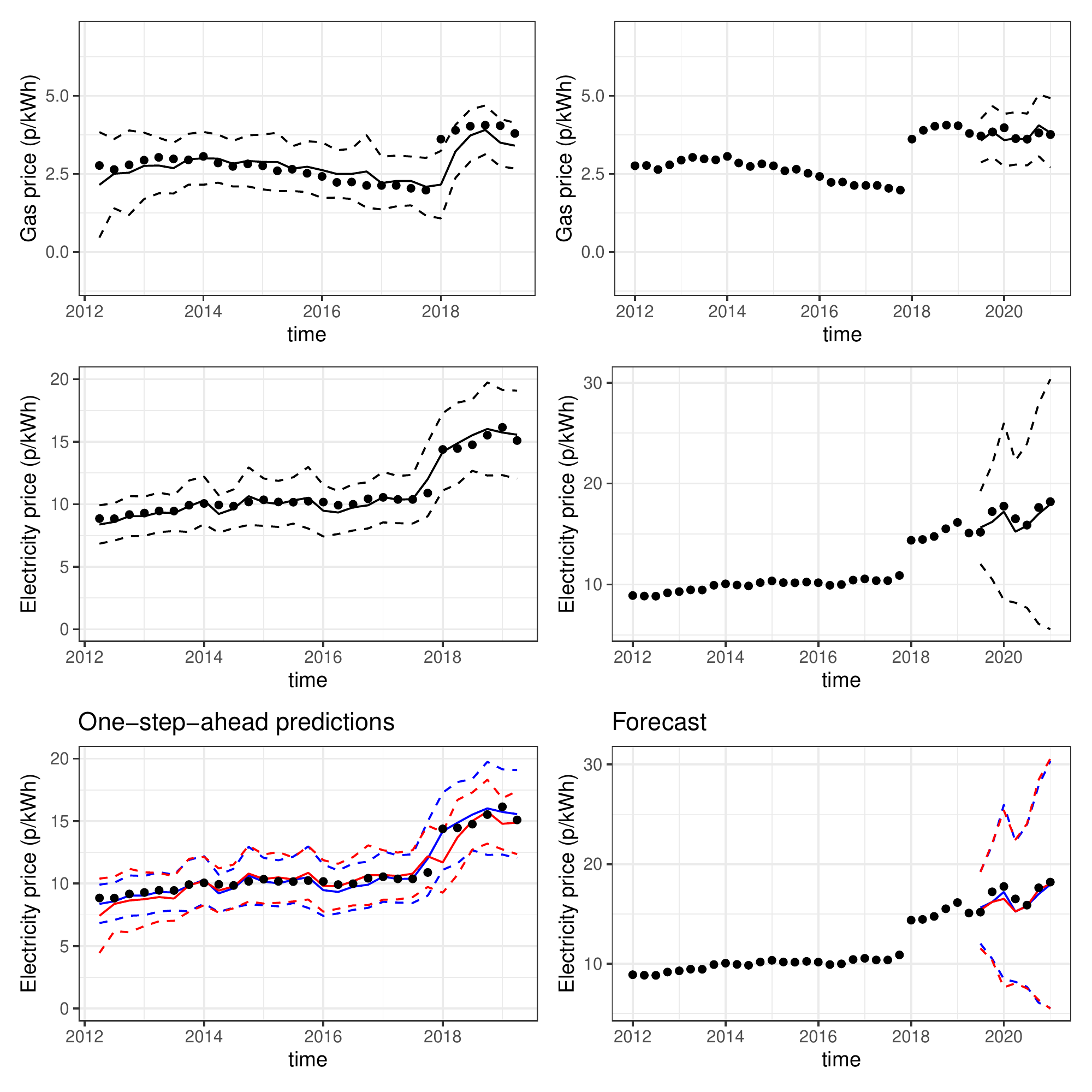}
\caption{Top row: dynamic linear regression fit for gas prices under Scenario 3. Middle row: dynamic linear regression fit for electricity prices under Scenario 3. Bottom row: LMDM fit (red) and DLM fit (blue) for electricity prices under Scenario 3. The solid line and dashed lines correspond to mean and two standard deviation prediction intervals. The points correspond to the observed prices.}\label{fig:Diagnostics_DLM_case3}
\end{center} 
\end{figure}
\end{supplement}
\end{document}